# Generation of Terahertz Radiation by Wave Mixing in Armchair Carbon Nanotubes

S.Y.Mensah[a], S. S. Abukari[a], N. G. Mensah[b], K. A. Dompreh[a], A. Twum[a] and F. K. A. Allotey[c]

[a]*Department of Physics, Laser and Fibre Optics Centre, University of Cape Coast, Cape Coast, Ghana*

[b]*Department of Mathematics, University of Cape Coast, Cape Coast, Ghana*

[c]*Institute of Mathematical Sciences, Accra, Ghana*

**Abstract**

Using semiclassical Boltzmann equation we have studied theoretically an effect of a direct current (DC) generation in undoped armchair carbon nanotube (CN) by mixing two coherent electromagnetic waves with commensurate frequencies i.e $\omega_1 = \Omega$ and $\omega_2 = 2\Omega$. We compared the results of the armchair with that of the zigzag carbon nanotubes for the same conditions ( i.e. when the normalized current is plotted against the amplitude). Quantitatively they agree with each other except that the absolute value of the peaks of the current for zigzag is about 1.1 times that of the armchair. We noticed that the current *j* is negative similar to that observed in zigzag CNs describing the same effect. However it is interesting to note that graph of normalized current against omega tau showed that the armchair is greater than that of the zigzag for about 1.1 times which is opposite. We also observed that when the phase shift θ lies between $\frac{\pi}{2}$ and $\frac{3\pi}{2}$ there is an inversion and the current becomes positive. We suggest the use of this approach for the generation of terahertz radiation and also for the determination of the relaxation time of electrons in carbon nanotubes.

*PACS codes*: 73.63.-b; 61.48.De

*Keywords*: Carbon nanotubes, harmonic wave mixing, direct current generation, terahertz radiation

## 1. Introduction

The problem of emission and reception of electromagnetic radiation has attracted the attention of the scientific community for a long time. Good sources of coherent electromagnetic radiation, its receivers and detectors exist for the radio-frequency, microwave and optical ranges of the spectrum. Many of these devices are based on semiconductor technology. Nowadays, the terahetz range (0.1- 1 THz, submillimitre wavelengths) is practically the last unexploited [1].

In 1971 Esaki and Tsu [2] and Romanov [3] suggested to use the superlattice (SSLs) as a nonlinear material for electromagnetic wave mixing and new harmonics generation. The theory of wave mixing in SSLs, based on a solution of the Boltzmann equation, with a constant relaxation time for single miniband have been studied in [4] and [5].

The first paper to study dc component of electric or magnetic field in a pure ac- driven circuits with nonlinear and symmetric characteristics can be found in [6] and a similar idea was also suggested by [7]. Soon after these theoretical predictions, the generation of the direct current (DC) due mixing of ac electric field and its second harmonic has been observed in [8] and independently by [9]. The effective inversion to the Bloch oscillations have been studied in [10] and this indicated a spontaneous creation of constant voltage and corresponding dc current that can be considered as a rectification of the THz in SSLs.

An enormous literature has evolved describing several nonlinear mechanisms that could be responsible for a generation of the DC at wave mixing in semiconductors [11-13]. Among the most important of them includes the heating mechanism where the nonlinearity is dependent on the relaxation constant on the electric field [13-16]. Independently, Goychuk and Hänggi [17] suggested another scheme of quantum rectification using a wave mixing of an ac field and its second harmonic in a single miniband of SSL. The approach of [17] is based on the theory of quantum ratchets and hence the necessary conditions for the appearance of DC include a dissipation (quantum noise) and extended periodic system.

In this work we study the effect of DC generation in armchair carbon nanotubes (CNs) due to a wave mixing of coherent electromagnetic radiations of commensurate frequencies. This effect is, in essence due to a nonparabolicity of the electron energy band and is stronger in systems like SSLs and CNs. We made a comparison of an effect of a direct current generation in a armchair CNs subjected to ac field and its second harmonic (n=2) with our earlier findings describing the same effect in zigzag CNs [18].

This work will be organised as follows: section 1 deals with introduction; in section 2, we establish the theory and solution of the problem; section 3, we discuss the results and draw conclusion.

## 2. Theory

Following the approach of [18] we consider an undoped single-wall armhair (n, n) carbon nanotubes (CNs) subjected to the electric mixing harmonic fields.

$$E(t) = E_1 \cos\omega_1 t + E_2 \cos(\omega_2 t + \theta) \quad (1)$$

We further consider the semiclassical approximation in which the motion of $\pi$-electrons are considered as classical motion of free quasi-particles in the field of crystalline lattice with dispersion law extracted from the quantum theory.

Considering the hexagonal crystalline structure of CNs and the tight binding approximation, the dispersion relation is given as

$$\varepsilon(s\Delta p_\varphi, p_z) \equiv \varepsilon_s(p_z) =$$

$$\pm \gamma_0 \left[1 + 4\cos(as\Delta p_\varphi)\cos\left(\frac{a}{\sqrt{3}}p_z\right) + 4\cos^2\left(\frac{a}{\sqrt{3}}p_z\right)\right]^{1/2} \quad (2)$$

for armchair CNs [19]

Where $\gamma_0 \sim 3.0 eV$ is the overlapping integral, $p_z$ is the axial component of quasimomentum, $\Delta p_\varphi$ is transverse quasimomentum level spacing and $s$ is an integer. The expression for $a$ in Eq (2) is given as

$$a = 3a_{c-c}/2\hbar \quad (3)$$

Where $a_{c-c} = 0.142 nm$ is the C-C bond length and $\hbar$ is Plank's constant divided by $2\pi$. The - and + signs correspond to the valence and conduction bands, respectively. Due to the transverse quantization of the quasi-momentum, its transverse component can take $n$ discrete values,

$$p_\varphi = s\Delta p_\varphi = \pi\sqrt{3}\, s/an \quad (s = 1 \ldots, n)$$

Unlike transverse quasimomentum $p_\varphi$, the axial quasimomentum $P_z$ is assumed to vary continuously within the range $0 \leq p_z \leq 2\pi/a$, which corresponds to the model of infinitely long CN $(L = \infty)$. This model is applicable to the case under consideration because we are restricted to temperatures and /or voltages well above the level spacing [19], ie. $k_B T \gg \varepsilon_C, \Delta\varepsilon$, where $k_B$ is Boltzmann constant, $T$ is the temperature, $\varepsilon_C$ is the charging energy. The energy level spacing $\Delta\varepsilon$ is given by

$$\Delta\varepsilon = \pi\hbar v_F/L \quad (4)$$

where $v_F$ is the Fermi speed and L is the carbon nanotube length [20].

Employing Boltzmann equation with a single relaxation time approximation.

$$\frac{\partial f(p)}{\partial t} + eE(t)\frac{\partial f(p)}{\partial P} = -\frac{[f(p) - f_0(p)]}{\tau} \quad (5)$$

Where e is the electron charge, $f_0(p)$ is the equilibrium distribution function , $f(p)$ is the distribution function, and $\tau$ is the relaxation time. The electric field $E(t)$ is applied along CNs axis. The relaxation term of Eq (5) describes the electron-phonon scattering [21, 22] electron-electron collisions, etc.

Expanding the distribution functions of interest in Fourier series as;

$$f_0(p) = \Delta p_\varphi \sum_{s=1}^{n} \delta(p_\varphi - s\Delta p_\varphi) \sum_{r=1}^{\prime} f_{rs}(t)\, e^{ibrp_z} \quad (6)$$

and

$$f(p,t) = \Delta p_\varphi \sum_{s=1}^{n} \delta(p_\varphi - s\Delta p_\varphi) \sum_{r \neq 1} f_{rs} e^{ibrp_z} \emptyset_v(t) \quad (7)$$

Where the coefficient, $\delta(x)$ is the Dirac delta function, $f_{rs}$ is the coefficient of the Fourier series and $\emptyset_v(t)$ is the factor by which the Fourier transform of the nonequilibrium distribution function differs from its equilibrium distribution counterpart.

$$f_{rs} = \frac{a}{2\pi \Delta p_\varphi S} \int_0^{\frac{2\pi}{a}} \frac{e^{-ibrp_z}}{1 + exp(\varepsilon_s(p_z)/k_B T)} dp_z \quad (8)$$

Where $b = \frac{a}{\sqrt{3}}$.

Substituting Eqs. (6) and (7) into Eq. (5), and solving with Eq. (1) we obtain

$$\emptyset_v(t) = \sum_{k_1,k_2=-\infty}^{\infty} \sum_{v_1,v_2=-\infty}^{\infty} J_{k_1}(\beta_1) J_{k_2}(\beta_2) J_{k_1+v_1}(\beta_1) J_{k_2+v_2}(\beta_2)$$

$$\times \left( \frac{(1 - i(k_1\omega_1 + k_2\omega_2)\tau)}{1 + ((k_1\omega_1 + k_2\omega_2)\tau)^2} \right) \times \{cos(v_1\omega_1 t + v_2(\omega_2 t + \theta))$$

$$- isin(v_1\omega_1 t + v_2(\omega_2 t + \theta))\} \quad (9)$$

where $\beta_1 = \frac{earE_1}{\omega_1}$, $\beta_2 = \frac{earE_2}{\omega_2}$, and $J_k(\beta)$ is the Bessel function of the $k^{th}$ order.

Similarly, expanding $\varepsilon_s(p_z)/\gamma_0$ in Fourier series with coefficients $\varepsilon_{rs}$

$$\frac{\varepsilon_s(p_z, s\Delta p_\varphi)}{\gamma_0} =$$

$$\varepsilon_s(p_z) = \sum_{r \neq 0} \varepsilon_{rs} e^{isbrp_z} \quad (10)$$

Where $\varepsilon_{rs} = \frac{a}{2\pi\gamma_0} \int_0^{\frac{2\pi}{a}} \varepsilon_s(p_z) e^{-isbrp_z} dp_z$ (11)

and expressing the velocity as

$$v_z(p_z, s\Delta p_\varphi) = \frac{\partial \varepsilon_s(p_z)}{\partial p_z} = \gamma_0 \sum_{r \neq 0} ibr \varepsilon_{rs} e^{isbrp_z} \quad (12)$$

We determine the surface current density a

$$j_z = \frac{2e}{(2\pi\hbar)^2} \iint f(P) v_z(P) d^2 p$$

or

$$j_z = \frac{2e}{(2\pi\hbar)^2} \sum_{s=1}^{n} \int_0^{\frac{2\pi}{a}} f(p_z, s\Delta p_\varphi) \emptyset_v(t) v_z(p_z, s\Delta p_\varphi) dp_z \quad (13)$$

and the integration is taken over the first Brillouin zone. Substituting Eqs. (7), (9) and (12) into (13), we obtain

$$j_z = \frac{8e\gamma_0}{\sqrt{3}\hbar n a_{c-c}} \sum_{r=1}^{\infty} r \left[ \frac{(k_1\omega_1 + k_2\omega_2)\tau cos(v_2\theta) + sin(v_2\theta)}{1 + ((k_1\omega_1 + k_2\omega_2)\tau)^2} \right]$$

$$\times \sum_{k_1,k_2=-\infty}^{\infty} \sum_{v_1,v_2=-\infty}^{\infty} J_{k_1}(\beta_1) J_{k_2}(\beta_2) J_{k_1-v_1}(\beta_1) J_{k_2-v_2}(\beta_2) \sum_{s=1}^{n} f_{rs} \varepsilon_{rs} \quad (14)$$

Linearizing this results with respect with respect to $E_2$ using

$$J_{\pm 1}(\beta_2) \sim \beta_2/2 \; ; \; J_0(\beta_2) \sim 1 - \left(\beta_2/4\right)$$

and then averaging the result with respect to time $t$, we obtain the direct current subjected to $\omega_1 = \Omega$ and $\omega_2 = 2\Omega$ as follows;

$$j_z = \frac{4e^2\gamma_0 a}{3\hbar n a_{c-c}} E_2 cos\theta \sum_{r=1}^{\infty} r^2 \sum_{k=-\infty}^{\infty} \frac{k J_k(\beta_1) J_{k-2}(\beta_1)}{1 + (k\Omega\tau)^2} \sum_{s=1}^{n} f_{rs} \varepsilon_{rs} \quad (15)$$

Subsequently $\Omega\tau$ will be represented by $z_c$.

### 3. Results, Discussion and Conclusion

Solving the Boltzmann equation with constant relaxation time τ, we obtain the exact expression for current density in armchair CNs subjected to an electric field with two frequencies $\omega_1 = \Omega$ and $\omega_2 = 2\Omega$ after lengthy analytical manipulation.

The current density $J_z$ is dependent on the electric field $E_2$ and $E_1$, the phase difference $\theta$, the frequency $\Omega$, the relaxation time $\tau$ and $n$ in a complex form as expressed in Eq. 15. We therefore sketched equation (15) using Matlab to understand the dynamics. Fig.1 represents the graph of $J_z/J_0$ on $\beta_1$ for $z_c = 0.3, 0.5, 0.9, 1$ and $2$. We observed that the current decreases rapidly, reaches a minimum value, $\beta_{min}$ and rises. For $z_c \ll 1$, the current density rises monotonously while for $z_c \geq 1$, the current rises and then oscillates. This indicates that at low frequency there is rectification while at high frequency some fluctuations occur. The rectification can be attributed to non ohmicity of the carbon nanotube for the situation where it Bloch oscillates. We also observed a shift of the $J_{min}$ to the left with increasing value of $z_c$.

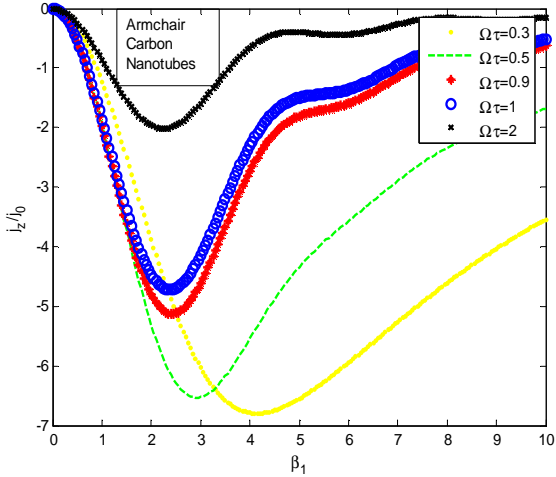

**Fig. 1.** $J_z/J_0$ is plotted against $\beta_1$ for $(-\cdot-)$ $zc = \Omega\tau = 0.3$; $(\cdots)$ $zc = \Omega\tau = 0.5$;

The behaviour of the current is similar to that observed in zigzag carbon nanotubes. In comparison with the result in zigzag carbon nanotubes for $z_c = 1$ the ratio $\frac{|J_{min}^{zigzag}|}{|J_{min}^{armchair}|} \approx 1.1$.

The result is also compared with superlattice. See Fig.2.

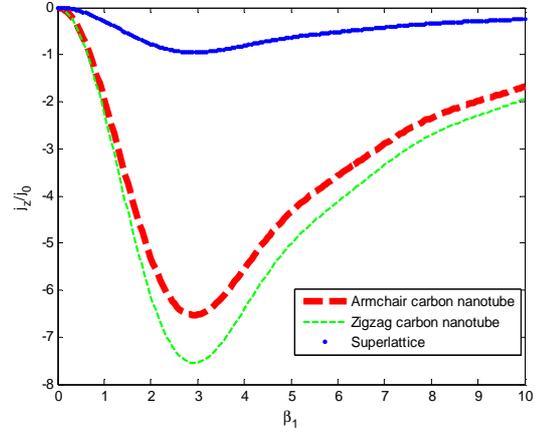

**Fig. 2.** $J_z/J_0$ is plotted against $\beta_1$ for $zc = \Omega\tau = 1$

We further studied the behavior of the current by sketching also the graph of $J_z/J_0$ against $z_c$ for $\beta_1 = 0.3, 0.5, 0.9, 1$ and $2$. The graph also displayed a negative differential conductivity. See Fig. 3.

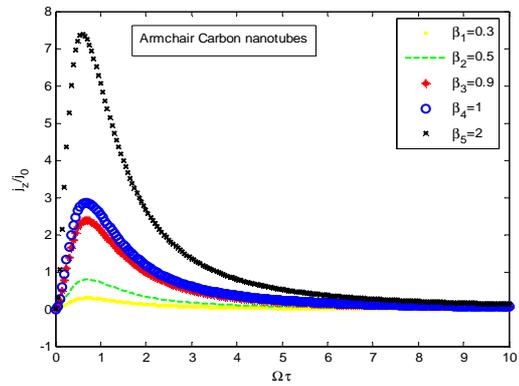

**Fig.3.** $J_z/J_0$ is plotted against $zc = \Omega\tau$ for $(-\cdot-)$ $\beta_1 = 0.3$; $(\cdots)$ $\beta_1 = 0.5$; $(***)$ $\beta_1 = 0.9$ ;$(\circ\circ\circ)$ $\beta_1 = 1$; $(\cdots)$ $\beta_1 = 2$.

Interestingly like in zigzag carbon nanotubes the current is always positive and has a maximum at the value $z_{c\ max} \approx 0.71$ irrespective of the amplitude of the electric field $E_1$. However, the peak of the current $j_{max}$ for armchair is greater than that of the zigzag by about 1.1. See Fig. 4.

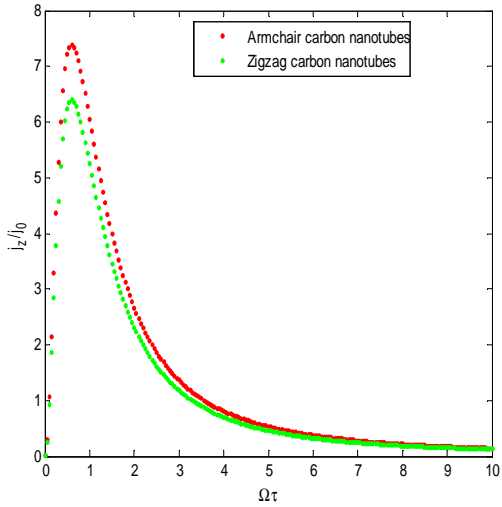

**Fig. 4.** $J_z/J_0$ is plotted against $zc = \Omega\tau$ for $\beta_1 = 1$

This is quite interesting because it is just opposite to the case when $j_z/j_0$ is sketched against $\beta_1$.

It is worthwhile to note that $z_{c\ max}$ can be used to determine the relaxation time of the electrons in the nanotube. e.g. $\tau \approx \frac{0.71}{\Omega}$ so knowing $\Omega$ you can determine $\tau$. On the other hand for typical value for $\tau$ of $10^{-13}s$ the frequency $\frac{\Omega}{2\pi}$ would be 1.2 THz.

Finally we sketched a 3 dimensional graph of the current against $\beta_1$ and $n$. for $\theta$ lying between $\frac{\pi}{2}$ and $\frac{3\pi}{2}$. An inversion was observed. See Fig. 5.

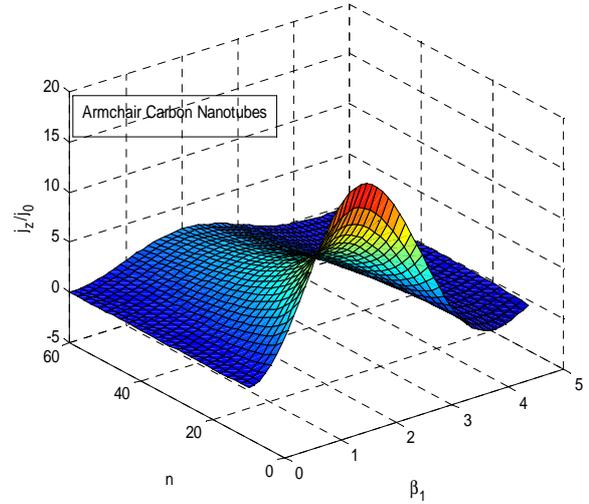

**Fig.5.** $J_z/J_0$ is plotted against $\beta_1$ for armchair carbon nanotubes.

In conclusion, we have studied the direct current generation due to the harmonic wave mixing in armchair carbon nanotubes and suggest the use of this approach in generation of THz radiation. The experimental conditions for an observation of the dc current effect are practically identical to those fulfilled in a recent experiment on the generation of harmonics of the THz radiation in a semiconductor superlattice [23-24]. This method can also be used to determine the relaxation time $\tau$.

**Reference**

[1] Kiril N. Alekseev and Feodor V. Kusmartsev arxiv: cond-mat/0012348v1, 2000

[2] Esaki L. and Tsu R., Appl. Phys. Lett., 19 (1971) 246.

[3] Romanov Yu. A., Optika i Spektr., 33 (1972) 917 (Sov. Phys. Optics and Spectroscopy).

[4] Bass F. G. and Tetervov A. P., Phys. Rep., 140 (1986) 237.

[5] Mensah S. Shmelev G. M. and Epshtein E.M., Izv. Vyzov, Fizika 6 (1988) 112

[6] Orlov L. K. and Romanov Yu. A., Fiz. Tverd. Tela 19 (1977) 726 (Sov Phys. Solid State); Romanov Yu. A., Orlov L. K.

[7] M. A. Rosenblat, Doklady AN SSR LXVIII, 497 (1949), in Russian.

[8] C. E. Skov and E. Pearlstein, "Sensitive method for the measurement of nonlinearity of electrical conduction", Review of Scientific Instruments 35, 962 (1964).

[9] Yu. K. Pozhela and H. J. Karlin, "Some remarks on microwave excitation of dc by hot electrons in germanium", Proc. IEEE 53, 1788 (1965).

[10] Alekseev K. N. et al. Phys. Rev. Lett., 80 (1998) 2669 [ also cond-mat/9709026 ]

[11] Patel C. K. N., Slusher R.E. and Fleury P.A., Phys Rev Lett., 17 (1966) 1011

[12] Wolf P.A. and Pearson G.A., Phys. Rev. Lett., 17 (1966) 1015

[13] Belyantsev A. M., Kozlov V. A. and Trifonov B.A, Phys. Status Sol. (b) 48 (1971) 581

[14] Fomin V. M. and Pokatilov E.P., Phys. Status Sol. (b) 97 (1980) 161

[15] Shmelev G, M., Tsurkan G. J. and Nguyen Xong Shon, Izv. Vyzov, Fizika 2(1985) 84 (Russian Physics Journal)

[16] Genkin V. N., Kozlov V. A. and Piskarev V.I. Fiz. Tekh. Polupr., 8 (1074) 2013 (Sov. Phys. Semicond)

[17] Goychuk I., and Hänggi P., Euophys. Lett., 43 (1998) 503

[18] Mensah, S.Y, Abukari, S. S, Mensah, N. G, Dompreh, K. A., Twum, A and Allotey, F. K. A. arxiv: cond-mat/1002.3233, 2010.

[19] Anton S. Maksimenko and Gregory Ya. Slepan, Physical Review Letters. Vol 84 No 2 (2000)

[20] Kane, C., Balents, L. and Fisher, M. P. A., Phys. Rev. Lett. 79, 5086 –5089, 1997.

[21] Lin, M. F. and Shung, K. W.K., Phys. Rev. B 52, pp. 8423–8438, 1995.

[22] Jishi, R. A., Dresselhaus, M. S., and Dresselhaus, G. Phys. Rev. B 48, 11385 - 11389, (1993).

[23] Kane, C. L., Mele, E. J., Lee, R. S., Fischer, J. E., Petit, P., Dai, H., Thess, A., Smalley, R. E.,. Verscheueren, A. R. M., Tans, S. J. and Dekker, C., Europhys. Lett. 41, 683-688 (1998).

[24] Winnerl, S. Schomburg, E. Brandl, S. Kus, O., Renk, K. F.. Wanke, M. C. Allen, S. J, Ignatov, A. A. Ustinov, V. Zhukov, A. and. Kop'ev, P. S."Frequency doubling and tripling of terahertz radiation in a GaAs/AlAs superlattice due to frequency modulation of Bloch oscillations", Appl. Phys. Lett. 77, 1259 (2000).